\documentclass[oldversion, letter]{aa} 
\usepackage{natbib}
\usepackage{amssymb}
\usepackage{graphicx}
\usepackage{txfonts}

\begin{document}
   \title{HST FUV C~{\sc iv} observations of the hot DG Tauri jet}

   \author{P. C. Schneider\inst{1}
          \and
          J. Eisl\"offel\inst{2}
          \and
          M. G\"udel\inst{3}
          \and
          H. M. G\"unther\inst{4}
          \and
          G. Herczeg\inst{5}
          \and
          J. Robrade\inst{1}
          \and
          J. H. M. M. Schmitt\inst{1}
          }

   \institute{Hamburger Sternwarte,
              Gojenbergsweg 112, 21029 Hamburg, Germany, \email{cschneider@hs.uni-hamburg.de}
              \and
              Th\"uringer Landessternwarte Tautenburg, Sternwarte 5, 07778 Tautenburg, Germany
              \and
              Universit\"at Wien, Dr.-Karl-Lueger-Ring 1, 1010 Wien, Austria
              \and
              Harvard-Smithsonian Center for Astrophysics, 60 Garden Street, Cambridge, MA 02138, USA
              \and
              The Kavli Institute for Astronomy and Astrophysics, Peking University, Yi He Yuan Lu 5, Hai Dian Qu, Beijing 100871, China
             }

   \date{Received .. / accepted ..}

  \abstract
   {
     
    Protostellar jets are tightly connected to the accretion process and regulate the angular momentum balance of accreting star-disk systems. The DG~Tau jet is one of the best-studied protostellar jets and contains plasma with temperatures ranging over three orders of magnitude within the innermost 50\,AU of the jet.
    We present new Hubble Space Telescope (HST) far ultraviolet (FUV) long-slit spectra spatially resolving the C~{\sc iv} emission ($T\sim10^5$\,K) from the jet for the first time, and quasi-simultaneous HST observations of optical forbidden emission lines ([O~{\sc i}], [N~{\sc ii}], [S~{\sc ii}] and [O~{\sc iii}]) and fluorescent $H_2$ lines. The C~{\sc iv} emission peaks at $\approx42$\,AU from the stellar position and has a FWHM of $\approx52$\,AU  along the jet. Its deprojected velocity of around 200\,km\,s$^{-1}$  decreases monotonically away from the driving source. 
    In addition, we compare our HST data with the X-ray emission from the DG~Tau jet. 
    We investigate the requirements to explain the data by an initially hot jet compared to local heating. Both scenarios indicate a mass loss by the $T\sim10^5$\,K jet of $\sim10^{-9}\,M_\odot$\,yr$^{-1}$, i.e., between the values for the lower temperature jet ($T\approx10^4\,K$) and the hotter X-ray emitting part ($T\gtrsim10^6$\,K). However, a simple initially hot wind requires a large launching region ($\sim1\,$AU), and we therefore favor local heating.
   }

   \maketitle


\section{Introduction \label{sect:intro}}
Outflow activity is a ubiquitous phenomenon of star formation and possibly of accretion processes in general. Jets can remove angular momentum and might thereby regulate the accretion process \citep[][and references therein]{Matt_2010}. During their early evolution, protostars are surrounded by a thick envelope and the inner parts of the outflows are usually hidden in optical and UV observations. When the envelope disperses, stars become visible in the optical as they enter the classical T Tauri stars (CTTS) phase. The outflows of CTTS can be observed very close to the central object. This allows us to investigate jet acceleration and collimation in great detail \citep[e.g.,][]{Coffey_2008}. Nevertheless, the driving mechanism of these outflows remains elusive; neither the acceleration nor the collimation of outflows is understood in detail. Magneto-centrifugally launched disk winds, with a possible stellar contribution, are currently the preferred model \citep[][]{Pelletier_1992,Ferreira_2006}. 

The CTTS DG~Tau \citep[$d\approx140$\,pc, ][]{Kenyon_1994}
shows complex outflow-related features with stationary and moving components \citep[the system parameters are summarized in][]{Guedel_2007}. The optical jet consists of different velocity components \citep[][]{Bacciotti_2000}. Individual emission regions (knots) at greater distances from DG~Tau exhibit clear proper motion \citep[][]{Eisloeffel_1998, Dougados_2000}, and \citet{Lavalley_Fouquet_2000} showed that the \   {optical} forbidden emission lines (FELs) are compatible with shock heating ($50\lesssim v_{shock} \lesssim 100$\,km\,s$^{-1}$).   Despite strong blueshifted C~{\sc iv} emission from the inner 1\arcsec (200\,AU) around DG~Tau seen in HST Goddard High-Resolution Spectrograph (GHRS) data \citep[][]{Ardila_2002}, previous HST Space Telescope Imaging Spectrograph (STIS) observations did not detect C~{\sc iv} emission. A non-stellar origin of the C~{\sc iv} emission might explain this finding because only the innermost 0\farcs1 (20\,AU) were covered by STIS \citep[][]{Herczeg_2006}.  DG~Tau is particularly interesting since it is the only CTTS with detected stationary jet X-ray emission. This X-ray emission is located at $d\approx0\farcs2$ (40\,AU) from the driving source \citep[][]{Schneider_2008,Guedel_2011}. In addition to the inner, stationary X-ray component, an outer X-ray emitting knot is present, which likely possesses proper motion \citep{Guedel_2005, Guedel_2008, Guedel_2011}. The jet of the class 0/I object L1551~IRS\,5 shows a similar X-ray morphology \citep[][]{Favata_2002,Schneider_2011}, which indicates that this phenomenon might not be exclusive to CTTS (or DG~Tau).
 
Here we present new HST STIS observations of the DG~Tau jet tracing the low temperature part ($T\approx10^3$\,K) with FUV $H_2$ emission, the $10^4$\,K gas with specific optical FELs such as [O~{\sc i}], and the $10^5$\,K plasma with FUV C~{\sc iv} and optical [O~{\sc iii}] emission. We also discuss the nature of the different temperature components within the innermost 150\,AU.

\section{Observations and data analysis}
Our long-slit observations consist of a FUV spectrum covering the C~{\sc iv} doublet, a medium-resolution optical spectrum of important diagnostic FELs, and a blue, low-resolution spectrum (Table~\ref{tab:obs}). 
DG~Tau was positioned  at the HST aimpoint, and
the $52\arcsec\times0\farcs2$ slit was oriented along the jet axis. The FUV observations have a velocity resolution of $\approx25$\,km\,s$^{-1}$ (FWHM) and a spatial resolution of $\lesssim0\farcs1$ (FWHM), judging from observations of point sources with the same setup. The resolution of the optical spectra is 0.9\,\AA (4.1\,\AA) and 0\farcs13 (0\farcs15) for the G750M  and G430L, respectively. All observations have nominal zero point accuracies better than $0\farcs03$ and 12\,km\,s$^{-1}$. Velocities are expressed in the stellar rest frame as measured from the Li $\lambda 6708$ line in our spectrum ($+32\pm2$~km\,s$^{-1})$.

\begin{figure}[t]
\centering

\includegraphics[width=0.48\textwidth, angle=00]{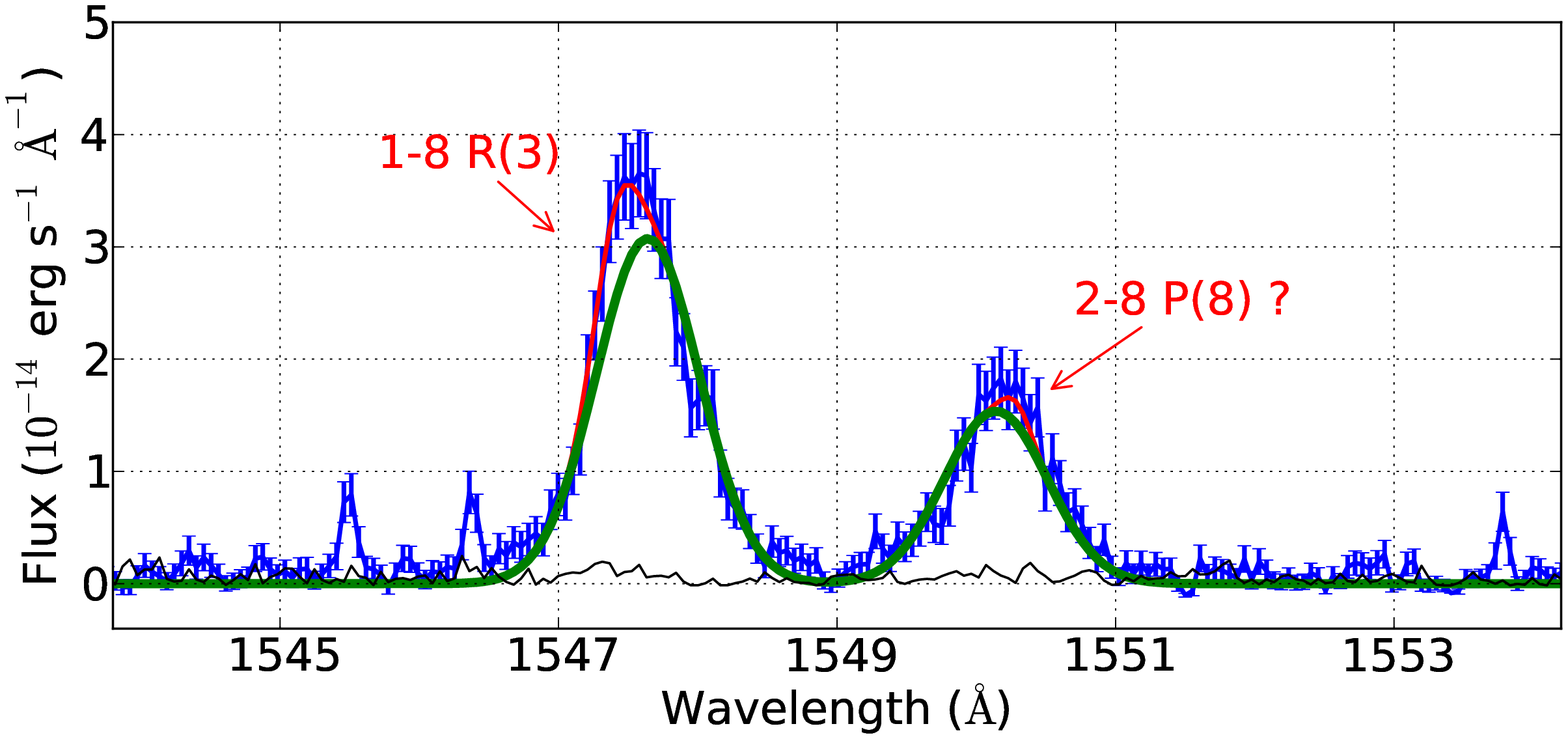}
\vspace*{-0.2cm}
\includegraphics[width=0.49\textwidth]{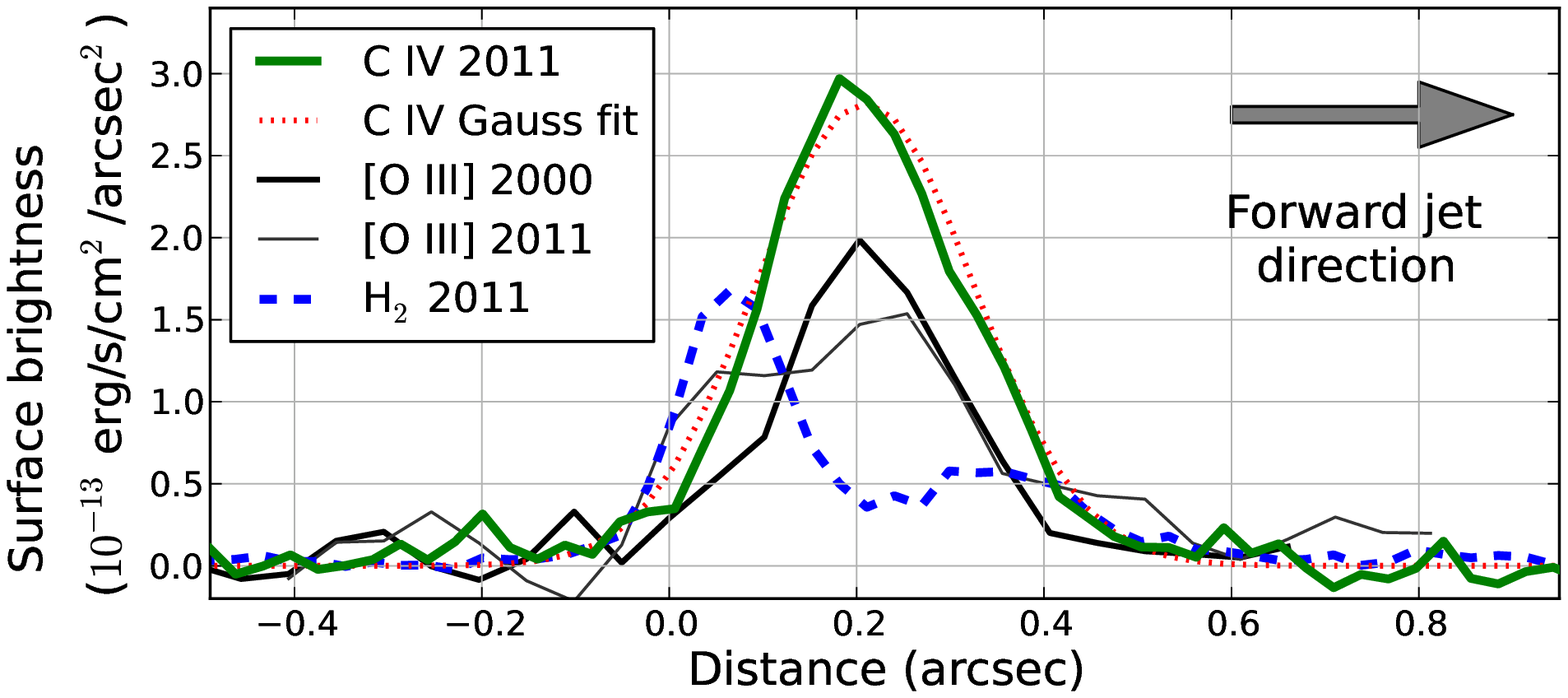}
\caption{
    {\bf Top}: Spectrum extracted between 0\farcs15\dots0\farcs3. The green line describes the Gaussian fit to the C~{\sc iv} emission, the red curve includes the $H_2$ contamination.
    {\bf Bottom}: C~{\sc iv} and $H_2$ flux along the jet axis summed between -400 to 0\,km\,s$^{-1}$ for the C~{\sc iv} emission and -80 to 0\,km\,s$^{-1}$ for the $H_2$ lines in our FUV spectrum. Included is [O~{\sc iii}] $\lambda5007$ emission seen in short G430L spectra.  \label{fig:both} \label{fig:specs} }\vspace*{-0.2cm}
\end{figure}

We used standard processing, i.e., the rectified flux-calibrated files and errors obtained by weighting the error instead of the variance for the FUV spectrum, which is more appropriate in the case of low-count statistics. The FUV Multi-Anode Microchannel Array
(MAMA) detector suffers from several hot pixels of varying amplitudes.
Visual inspection of the region around the C~{\sc iv} doublet shows that about ten hot pixel islands are expected in the co-added spectrum within 1543.4\,\AA{}\dots 1554.0\,\AA{} (-1000 to +600\,km\,s$^{-1}$ around the C~{\sc iv} doublet) and $1\farcs2$ to $-0\farcs5$. Their flux is $<4\,\%$ of the total flux in this region. There is virtually no stellar light in the FUV spectra, but the C~{\sc iv} $\lambda$1548.2\,\AA{} line is contaminated by $H_2$ emission (1-8 R(3) $\lambda\,1547.3$\,\AA); C~{\sc iv} $\lambda$1550.7\,\AA{}  is only  marginally contaminated by $H_2$ 2-8 P(8) $\lambda\,1550.3$\,\AA{} as no line within the same fluorescence route is significantly detected (contamination $\lesssim5\,$\%). The $H_2$ contamination is about 13\% of the total emission. In order to remove the $H_2$ contamination of the C~{\sc iv} emission, we subtracted the distribution of the strongest $H_2$ line in the same fluorescence route of the contaminating line (1-8 P(5) at 1562.4\,\AA) scaled by the theoretical line ratio.
We removed the stellar continuum from the optical position velocity diagrams (PVDs) by measuring its intensity at line-free regions at both sides of the emission line and interpolating over the region of interest. Sampling effects caused by the tilt of the spectral trace with respect to the detector limit the accuracy of this procedure. 

The  inner jet's  X-ray absorption is consistent with the interstellar absorption towards the Taurus star-forming region, and we translate the X-ray absorption \citep[$N_H =1.1\times10^{21}$\,cm$^{-2}$,][]{Guedel_2008} to $A_V = 0.55$ (E(B-V) = 0.18) using the standard conversion \citep{Vuong_2003}.  This leads to a transmission of 27\,\%  through the line of sight in the FUV. DG Tau itself is more strongly absorbed, both in X-rays and in the UV/optical \citep{Guedel_2007, Gullbring_2000}, which allowed us to separate stellar emission from jet emission in the X-ray domain.
We assume a jet inclination angle of $i\approx42^\circ$. Given errors are 90\% statistical errors and do not include the effect of the spatial distribution within the slit, quoted velocities are projected ones.

\vspace*{-0.2cm}
\section{Results}
Figure~\ref{fig:specs} (top) shows the spectrum obtained around the peak of the C~{\sc iv} emission. The velocity-integrated spatial distributions of C~{\sc iv}, $H_2$, and [O~{\sc iii}] emission along the jet axis are shown in Fig.~\ref{fig:both} (bottom). 
The C~{\sc iv} PVD is shown in  Fig.~\ref{fig:pvds}; both C~{\sc iv} lines were co-added since they are compatible with the theoretical line ratio of two for optically thin emission. 

\subsection{Positions, sizes, and velocities}
The velocity-integrated C~{\sc iv} flux peaks at $0\farcs2$ along the forward jet direction from DG~Tau (deprojected 42\,AU). 
The deconvolved Gaussian FWHM of the velocity-integrated C~{\sc iv} emission is 0\farcs25 (52\,AU). Independent of the assumed flux distribution intrinsic C~{\sc iv} emission must be present at $0\farcs05 - 0\farcs1$ (10-21\,AU) and extend beyond $0\farcs35$ (73\,AU). 
The spatial resolution of the X-ray observations (pixel size 0\farcs5) does not allow us a morphological comparison with hotter X-ray emitting plasma ($T\gtrsim3\times10^6$\,K) but the centroid of the X-ray jet is located 0\farcs14 to 0\farcs21 from DG Tau, depending on the method and observations used (the range also corresponds roughly to the statistical uncertainty). This places the C~{\sc iv} peak slightly ($<$13\,AU) further away from DG~Tau than the jet X-ray emission.

\begin{table}[t!]
\begin{minipage}[h]{0.5\textwidth}
\renewcommand{\footnoterule}{}
  \caption{Analyzed HST STIS observations \label{tab:obs} }
  \begin{tabular}{c c c c c c} \hline \hline
  Obs.-date  & Grating & Exp. & Wavelength & arcsec & $\AA$\\
             &                   & time & coverage (\AA) & pix$^{-1}$ & pix$^{-1}$\\
  \hline
  2011-02-17  & G140M & 14\,ks & 1514 - 1567 & 0.03 & 0.05 \\
  2011-02-17  &  G430L & 60\,s & 2957 - 5703 & 0.05 & 2.75\\
  2000-10-22  &  G430L & 240\,s & 2957 - 5703 & 0.05 & 2.75\\
  2011-02-14  & G750M & 2.3\,ks & 6295 - 6863 & 0.05 & 0.55\\
  \hline
  \end{tabular}
  \end{minipage}
\vspace*{-0.3cm}  
\end{table}

\begin{figure*}[t]
\includegraphics[width=0.25\textwidth, angle=00]{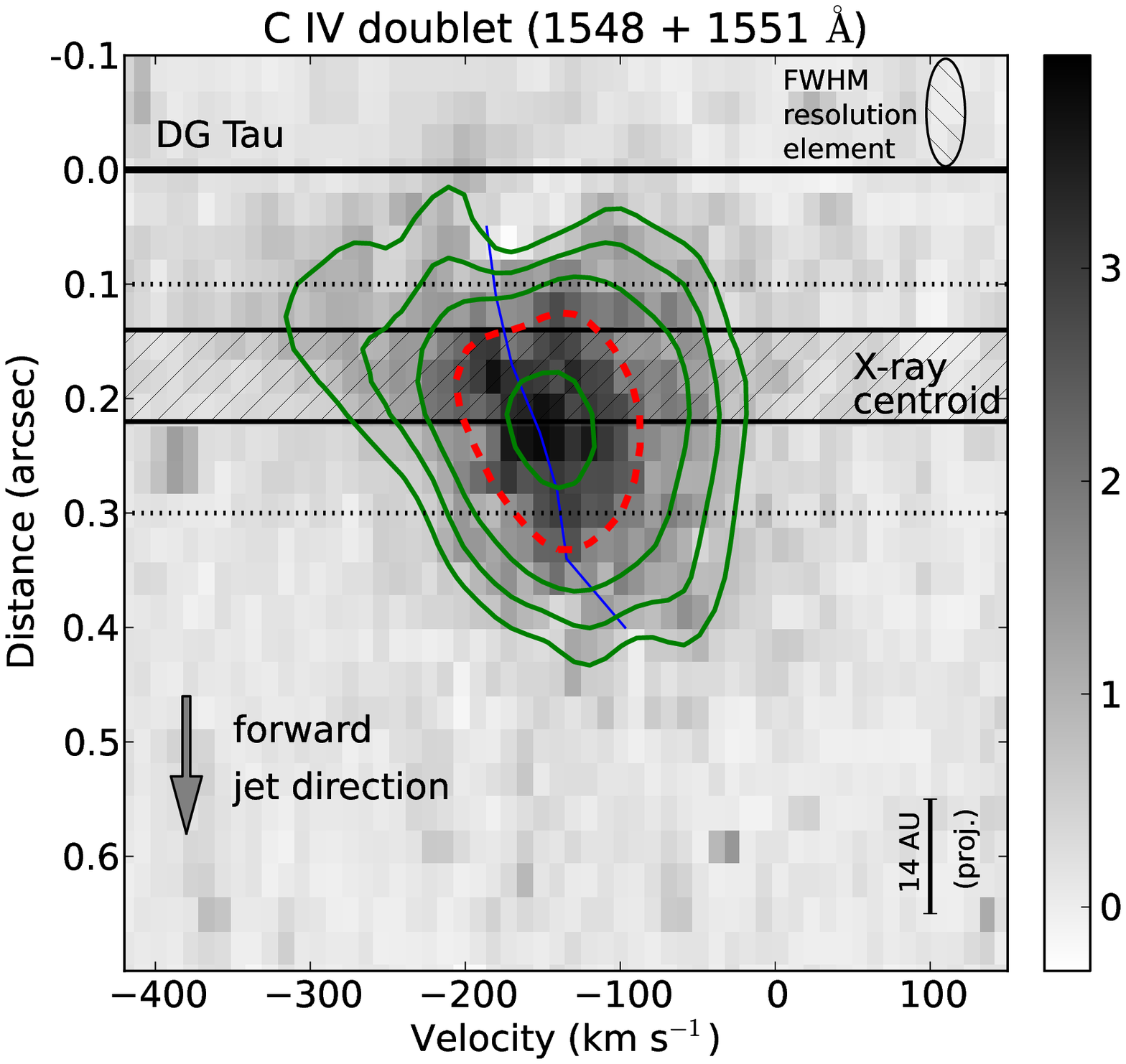}
\hspace*{-0.1cm} 
\includegraphics[height=0.24\textwidth, angle=0]{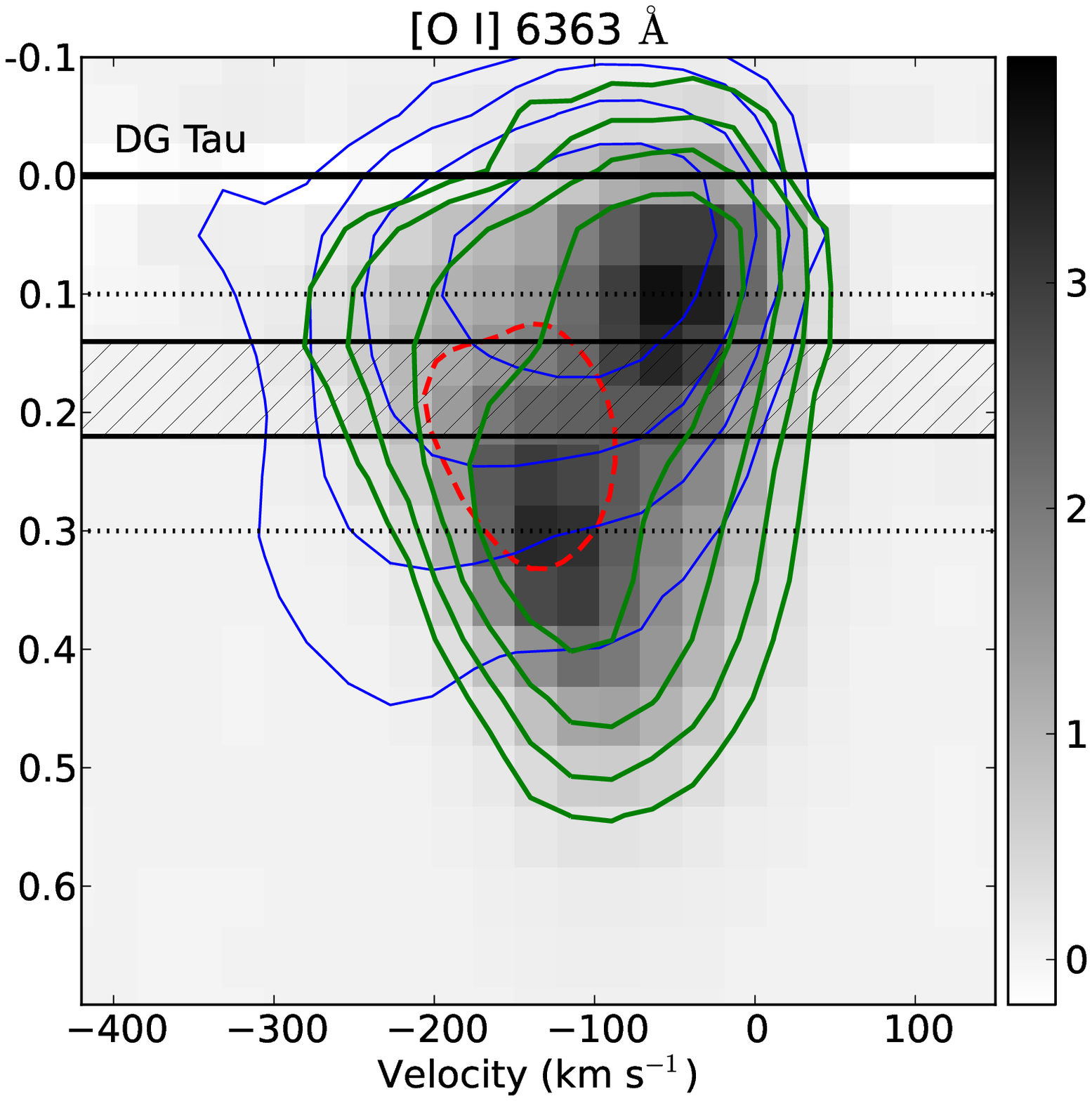}
\includegraphics[height=0.24\textwidth]{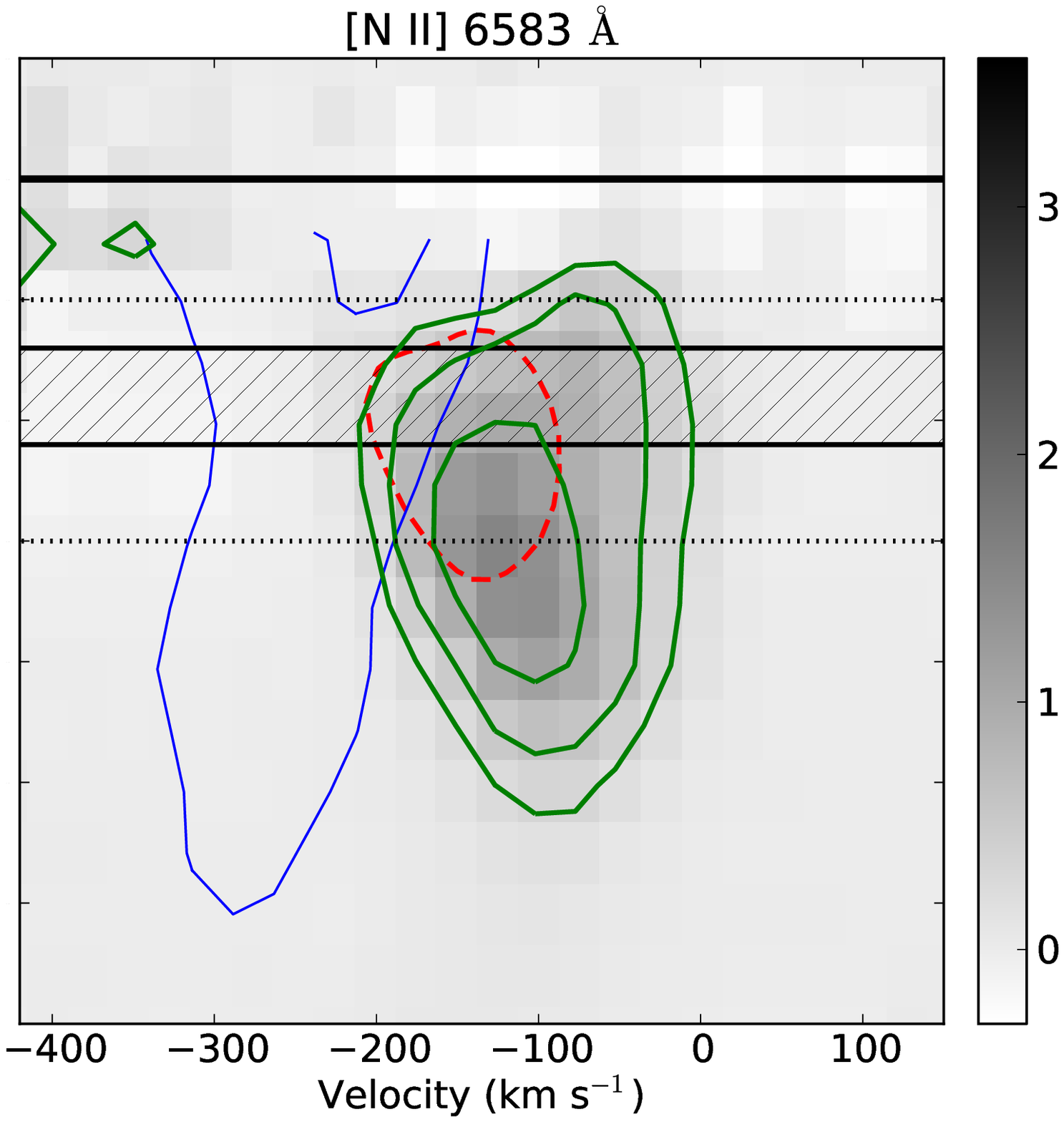} 
\includegraphics[height=0.24\textwidth]{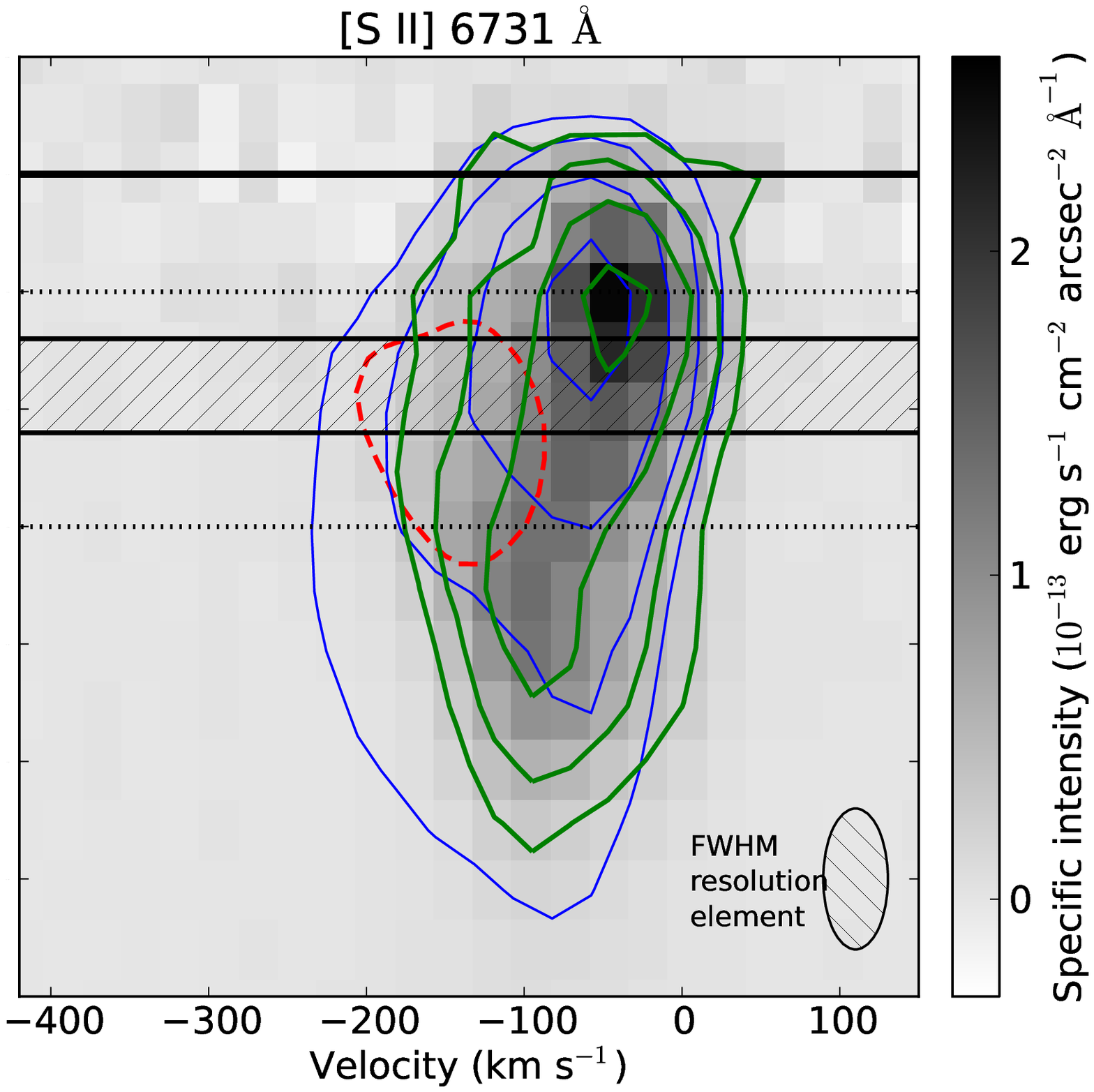}
\caption{PVDs of important diagnostic lines with associated contours. The red dashed contour indicates the C~{\sc iv} emission and the horizontal dotted lines give the peak location of the two optical knots. The shaded area indicates the centroid of the inner X-ray jet (spatial extent: 
$\sim 0\farcs1$), 
the blue line visualizes the velocity of the C~{\sc iv} emission, and the blue contours pertain to the central jet emission in the 1999 STIS data. Contours   start at $7\times10^{-14}$\,erg\,s$^{-1}$\,cm$^{-2}$\,\AA$^{-1}$ (increase by $\sqrt{2}$) and  at $2.5\times10^{-14}$\,erg\,s$^{-1}$\,cm$^{-2}$\,\AA$^{-1}$ (increase by a factor of 2) for  C~{\sc iv} and optical emission, respectively. }
\label{fig:pvds} 
\end{figure*} 

The velocity and FWHM of the C~{\sc iv} emission are largest close to DG~Tau and continuously decrease with increasing distance (Fig.~\ref{fig:pvds}). From fitting Gaussians to the C~{\sc iv} emission, we find that the mean velocity drops from  $\approx-180$\,km\,s$^{-1}$ at $0\farcs1$ to  $\approx-100$\,km\,s$^{-1}$ at $0\farcs4$. The FWHM drops from above $\approx200$\,km\,s$^{-1}$ to $\approx130$\,km\,s$^{-1}$ over the same region.
We do not detect significant C~{\sc iv} emission at greater distances from DG~Tau, although the STIS spectrum covers the outer X-ray emitting knot. The more strongly absorbed counter jet is also not visible in the FUV. However, both components are visible in the optical spectrum (Schneider et~al., in prep.).
The C~{\sc iv} properties are consistent with the non-detection in the less sensitive archival STIS E140M spectrum of DG~Tau \citep[][]{Herczeg_2006} because that observation covers only the inner $0\farcs1$ of the jet. 

The [O~{\sc iii}] $\lambda5007$ emission seen in the short G430L is statistically compatible with the C~{\sc iv} emission. Therefore, we expect that there is no strong change in circumstellar absorption in the region where C~{\sc iv} emission is observed, because the ratio of FUV to optical transmission changes with increasing $A_V$.
Another short G430L spectrum  of DG~Tau from the year 2000 with the slit almost aligned along the jet (PA=230$^\circ$) shows [O~{\sc iii}] emission with a similiar spatial distribution, but more blueshifted ($-249_{-45}^{+27}$ vs. $-88^{+119}_{-91}$\,km\,s$^{-1}$), i.e., comparable with the C~{\sc iv} emission seen with GHRS. The location of [O~{\sc iii}] emission at $\approx0\farcs2$ in two observations and the stationarity of the X-ray emission suggest that the position of the C~{\sc iv} emission, tracing similar temperatures as O~{\sc iii}, is likely also stationary.  

The peak of the C~{\sc iv} emission corresponds to a local minimum of the $H_2$ emission (Fig.~\ref{fig:both}). The inner $H_2$ emission at $0\farcs1$  is slightly faster ($30\pm4$\,km\,s$^{-1}$)  than the outer component at $0\farcs3$ ($20\pm6$\,km\,s$^{-1}$), with no significant difference in their mean FWHM of about $43\pm8$\,km\,s$^{-1}$. 

The low-temperature optical FELs [O~{\sc i}], [N~{\sc ii}], and [S~{\sc ii}] resemble the two $H_2$ components at considerably higher velocities; their PVDs (Fig.~\ref{fig:pvds}) show a  low-velocity component (LVC, v$\approx60$\,km\,s$^{-1}$) at $0\farcs1$ and a medium-velocity component (MVC, v$\approx130$\,km\,s$^{-1}$) at $0\farcs3$. This structure would appear like an acceleration if both components were unresolved, but velocity and FWHM of the MVC decrease beyond $0\farcs3$. 
The C~{\sc iv} emission is located mainly between the LVC and MVC, but with higher velocities.
The central-slit data of the 1999 STIS observation \citep[slit-width 0\farcs1, ][]{Bacciotti_2000} show low-velocity material at $\approx0\farcs1$ from DG~Tau, i.e., at the same position as the 2011 LVC. Additionally, the 1999 STIS data show FEL emission at $0\farcs1$ with velocities corresponding roughly to those of the 2011 C~{\sc iv} emission.  Assuming that the higher velocity material seen 1999 in optical FELs represents the same  transient part of the jet as the 2011 MVC, it is reasonable to attribute the inner LVC at $\approx0\farcs1$ to stationary emission, as suggested by \citet[][]{Lavalley_1997}, who compared two epochs of ground-based observations.  

\vspace*{-0.2cm}
\subsection{Fluxes, emission measure, and densities}
The total C~{\sc iv} flux is $(2.0 \pm 0.1) \times 10^{-14}$\,erg\,s$^{-1}$\,cm$^{-2}$ (1548\,\AA{} + 1551\,\AA).  The [O~{\sc iii}]\,$\lambda5007$ flux is $1.9_{-0.6}^{+1.4}\times10^{-14}$\,erg\,s$^{-1}$\,cm$^{-2}$. For the adapted $A_V$ (see sect.~\ref{sect:intro}), the dust extinction corrected luminosities are $L_{\mbox{C {\sc iv}}} \approx (1.8\pm0.2)\times10^{29} \mbox{erg\,s}^{-1}$ and $L_{\mbox{[O {\sc iii}]}}\approx(8_{-0.3}^{+0.6})\times10^{28} \mbox{erg\,s}^{-1}$. Compared to the X-ray emission from the jet ($L_X\approx10^{29}$\,erg\,s$^{-1}$\,cm$^{-2}$), these values are of the same magnitude, but also show that more energy is emitted by the $\sim10^5\,$K than by the $T\gtrsim3\times10^6$\,K plasma. The flux in our spectrum is about half the flux seen in the 1996 GHRS spectrum \citep[$(4.4 \pm 0.4) \times 10^{-14}$\,erg\,s$^{-1}$\,cm$^{-2}$,][]{Ardila_2002},  but that spectrum also shows C~{\sc iv} emission mainly at higher velocities ($v\sim-260$\,km\,s$^{-1}$) and thus deviates significantly from our data. 

The C~{\sc iv} emission probably traces a distribution of temperatures, but the emissivity strongly peaks around $10^5$\,K. Using the \citet{Mazzotta_1998} ionization equilibrium and a C-abundance of $3\times10^{-4}$, we find a maximum radiative loss function of $6.8\times10^{-23}$\,erg\,s$^{-1}$\,cm$^3$, so that the
emission measure 
\begin{equation}
EM = L_{\mbox{C~{\sc iv}}} \, \Lambda_{\mbox{C~{\sc iv}}}^{-1}   \approx 2.6\times10^{51}\,\mbox{cm}^{-3} 
\end{equation}
can be regarded as a lower limit on the required $EM$. Consequently, the density of the plasma assuming a uniformly emitting cylinder and a volume-filling factor $f$ 
\begin{equation}
n_e = \left(\frac{EM}{f V} \right)^{0.5} = 8\times10^3 \left(\frac{1}{f} \right)^{0.5} \left( \frac{40\,\mbox{AU}}{l}\right)^{0.5} \left( \frac{10\,\mbox{AU}}{r}\right) \mbox{cm}^{-3}
\end{equation}
is also a lower limit ($r$ and $l$ are radius and length).
The density sensitive [S~{\sc ii}] $\lambda\lambda$ 6716/6731 lines are in the critical density limit ($n_{crit}=2\times10^4$\,cm$^{-3}$) within the innermost $0\farcs4$ (79\,AU) of the jet,  i.e., at the positions of the inner knots. This density limit is comparable to that of the $10^5\,$K plasma, assuming a filling factor of unity (Eq.~2), but the [S~{\sc ii}] emission traces lower flowspeeds and temperatures than the C~{\sc iv} emission. On the other hand, the [O~{\sc i}] $\lambda$5577 to $\lambda$6300 ratio of 0.2 at 0\farcs1 sets a strict density limit of $n<10^{10}\,$cm$^{-3}$ \citep[cf. Fig.~12 in][]{Hartigan_1995}.

\section{The nature of the C~{\sc iv} emission}
There are two possibilities for the existence of plasma exceeding $T\gtrsim10^5$\,K within the jet  at $\approx40\,$AU from DG~Tau, i.e., the C~{\sc iv} and X-ray emitting material: either some part of the outflow is already hot close to the launching point or the jet is sufficiently heated while flowing outwards.

\subsection{A hot inner outflow}
First, we consider a cylindrical/conical outflow that is launched with a sufficient temperature to explain the observed spatial offsets as a result of cooling and a special absorption geometry; how such an outflow might be confined and stabilized is beyond the scope of this letter.  The C~{\sc iv} emitting plasma has a short radiative cooling time, e.g., about one day corresponding to a length of 0.1\,AU along the jet axis at $v=200\,$km\,s$^{-1}$ for $n_e=10^6$\,cm$^{-3}$ ($\tau_{rad}\sim n_e^{-1}$). Therefore, the C~{\sc iv} emitting region must have a large extent perpendicular to the jet axis to provide the required $EM$. Using Eq.~(2) with $l=l_{cool}=0.1\,$AU$/n_6$ and assuming that C~{\sc iv} is emitted between $\log T=4.9$ and $5.2$, a filling factor of unity ($f=1$), and a cylindrical emission region, we estimate that the radius
must be $r=1.6\,$AU\,$n_6^{-0.5}$ with $n_6$ density in $10^6\,$cm$^{-3}$ ($n_e \approx n$ at the temperatures at hand).  Such a wind has $\dot{M}=v \mu m_H n \pi r^2 \approx10^{-9} M_\odot\,\mbox{yr}^{-1}$.
Densities around $10^4$\,cm$^{-3}$ give a cooling distance comparable to the observed offset for $T_0$ slightly above $10^5\,$K, and the peak of the C~{\sc iv} emission might simply denote the location where the C~{\sc iv} radiative losses are largest ($T=10^5\,$K). However, this requires that the disk launches a wind with $T>10^5\,$K over $r>10\,$AU and does not explain the proximity of X-ray and C~{\sc iv} emission. These issues can be addressed by proposing higher initial temperatures and densities, e.g.,  $T=3\times10^6$\,K and $n_e=10^6$\,cm$^{-3}$. In this case, the X-ray emission close to DG~Tau must be absorbed to explain the offset of the X-ray emission and the low X-ray to C~{\sc iv} ratio, since such a radiative wind has an X-ray to C~{\sc iv} emission ratio of $\sim40$  \citep[estimated using the CHIANTI database, ][]{Dere_2009}. We note that the X-ray absorption of DG~Tau reduces the soft X-ray transmission by a sufficient factor ($\sim10^2$). On the other hand, the X-ray data exclude initial temperatures exceeding a few $10^7\,$K, which would allow $n\gtrsim10^7$\,cm$^{-3}$. Therefore, the launching radius is $r\gtrsim0.5$\,AU. The observed distribution of C~{\sc iv} emission along the jet might result from different densities and velocities of individual plasmoids \citep[cf.][]{Skinner_2011}, whose cooling times might be altered by thermal conduction. 

The expansion of a conical wind adds an extra cooling term and thus requires higher initial temperatures. The emission measure per unit length along the jet axis is also reduced compared to a cylindrical outflow. As a result, higher initial densities are required to produce a similar C~{\sc iv} luminosity compared to a radiative wind with the same intial radius. Therefore, a conical outflow does not allow smaller launching radii. We thus regard an outflow with an initial temperature well above $10^6$\,K launched over $r\gtrsim0.5\,$AU unlikely and conclude that the heating must happen further out.

\subsection{Local heating}
Another possibility is that shocks locally heat the outflowing gas. Magnetic fields reduce the effective shock velocity and increase the cooling distance by limiting the compression of the post-shock plasma \citep{Hartigan_2007}.  For negligible magnetic fields, however, the cooling distance is very short \citep[less than one-tenth of an AU for $n_e=10^6$\,cm$^{-3}$ and $v_{shock}=200\,$km\,s$^{-1}$, ][]{Hartigan_1987}.   Therefore, the observed length of the emission region along the jet axis results either from a single large oblique shock and inclination effects or from a number of small  shocks at different streamlines since the observed velocities argue against a number of shocks for a single streamline.

Using the shock code of \citet{Guenther_2007} for negligible magnetic fields perpendicular to the shock surface, the detected (dereddened) soft X-ray luminosity of \hbox{$\approx10^{29}$\,erg\,s$^{-1}$} is insufficient to explain the C~{\sc iv} emission as cooled, previously X-ray-emitting plasma (a stationary 1D shock with $v_{shock}\approx 400$\,km\,s$^{-1}$  radiates less than 20\% of its soft X-ray luminosity in C~{\sc iv}). Without invoking extra-absorption for the soft X-ray emission, the observed C~{\sc iv} emission requires additional heating, e.g.,  additional shocks with lower velocities,  than the very high-velocity shocks responsible for the X-ray emission.  

For the shock scenario, we can estimate the required mass-loss from the fraction of the post-shock gas seen in C~{\sc iv} \citep[cf. estimate for O~{\sc i} emission in][]{Hartigan_1995}. For a stationary oblique shock, the ratios of shock to jet velocity and shock to jet surface are equal and cancel out in the calculation. From the \citet{Hartigan_1987} models, we estimate that every atom crossing the shock emits $N\approx0.3$ C~{\sc iv} photons, so that
$$
\dot{M}= \frac{L_{{C IV}}}{N h\nu_{C IV}} \mu m_H
\approx 2\times10^{-9}\left(\frac{0.3}{N}\right) \left(\frac{L_{C IV}}{1.8\times10^{29}\,\mbox{erg}\mbox{s}^{-1}}\right)\,M_\odot\,\mbox{yr}^{-1}\,.
$$
This is lower than the values derived from near-IR jet tracers \citep[$\dot{M} \gtrsim 10^{-8} M_\odot$\,yr$^{-1}$][]{Agra_Amboage_2011}. Possibly only a fraction of the outflow is heated to C~{\sc iv} emitting temperatures.

Alternatively, the jet might be locally heated by magnetic reconnection. 
We utilize the fact that the magnetic field is frozen into the plasma in a sufficiently ionized gas, i.e., is co-moving with the plasma. For an order of magnitude estimate, we approximate the inflow of magnetic energy by 
$$
L_{mag} \sim v_B A_r \frac{B^2}{8\pi} \sim 10^{32}\mbox{erg\,s}^{-1} \left(\frac{v_B}{200\,\mbox{km}\,\mbox{s}^{-1} }\right) \left(\frac{r_A}{10\,\mbox{AU}} \right)^2 \left( \frac{B}{50\,\mbox{mG}} \right)^2 \nonumber , 
$$
where $v_B$ describes the velocity of the plasma component carrying the magnetic field through the area $A_r$ (radius $r_A$) into the reconnection region, and we used the \citet{Hartigan_2007} estimate of $B$ at 50\,AU. A luminosity of $L_{mag}\gtrsim3\times10^{30}$\,erg\,s$^{-1}$ is required to power the C~{\sc iv} emitting plasma since its total radiative loss is about 15\,times $L_{C IV}$. Although probably only a small fraction of $L_{mag}$ is used for heating, it appears possible that magnetic heating is sufficient to heat the high-temperature plasma.  

We conclude that local heating can explain the observed high-temperature plasma within the DG~Tau jet. The apparent stationarity of the high-temperature plasma then requires a quasi stationary process at some distance to DG~Tau. The location of the high-temperature plasma in the collimation region makes an association with oblique shocks due to collimation likely. Possibly the highest velocity shocks are closer to the jet axis and to the source, while slower shocks exist at larger distances which explains the offset between X-ray and C~{\sc iv} emission.

\begin{acknowledgements}
PCS is supported by the DLR under grant 50OR1112.
HMG is supported by NASA through Chandra Award Number GO1-12067X on behalf of NASA under contract NAS8-03060.
This paper is based on observations made with the NASA/ESA Hubble Space Telescope.
\end{acknowledgements}
\bibliographystyle{aa}
\bibliography{dg}
\onecolumn
\end{document}